\documentclass[%
 reprint,
 amsmath,amssymb,
 aps,
]{revtex4-2}

\usepackage{graphicx}
\usepackage{dcolumn}
\usepackage{bm}
\usepackage[colorlinks=true]{hyperref}

\begin{document}

\preprint{APS/123-QED}

\title{Forget metamaterial: It does not improve sound absorption performance as it claims}

\author{Chao~Shen}
 \altaffiliation[Also at ]{Department of Mechanical Engineering, The University of Hong Kong, Hong Kong Special Administrative Region, China}
\author{Yu~Liu}%
 \email{liuy@sustech.edu.cn}
 
\affiliation{%
 Department of Mechanics and Aerospace Engineering, Southern University of Science and Technology, Shenzhen 518055, China}%

\author{Tianquan Tang}
\author{Lixi~Huang}
\affiliation{%
 Department of Mechanical Engineering, The University of Hong Kong, Hong Kong Special Administrative Region, China
}%

\date{\today}

\begin{abstract}
The term `sub-wavelength' is commonly used to describe innovative sound-absorbing structures usually labeled as `metamaterials'. Such structures, however, inherently do not bring groundbreaking advancements. This study addresses the limitations imposed by the thickness criterion of Yang et al.~\cite{yang2017optimal} by introducing the concept of equivalent mass-spring-damping parameters within the resonator framework. This innovative approach introduces an index of `half-absorption bandwidth' to effectively overcome the thickness restriction.
Four practical cases are then presented to correct prevalent misleading conceptions about low-frequency, broadband absorption as claimed. The phenomenon of mass disappearing in the expression of sound absorption coefficient supports the conclusion that volume is the only determinant factor in sound absorption performance. Any attempts to improve sound absorption solely through geometry and structural designs would inevitably sacrifice the half-absorption bandwidth. Additionally, the concept of negative stiffness or bulk modulus is merely a mathematical convention without any real improvement in absorption performance.
Overall, this research focuses on the physical mechanism of sound-absorbing structures by correcting traditional misunderstandings, and offers a comprehensive framework for assessing and enhancing sound absorption.

\end{abstract}

\maketitle


\textit{Introduction.}---Sound absorption using resonators is a complex and fascinating area of research. Broadband sound absorption in low-frequency range is particularly challenging due to the long wavelength~\cite{yang2017optimal,ryoo2022broadband}. Many researchers employed coiling configurations~\cite{li2016acoustic,gao2021hybrid,wu2019low,LiuChenkai2021,liang2012extreme,long2019broadband,landi2018acoustic,cheng2015ultra}, elongated necks~\cite{huang2020compact,huang2019acoustic,duan2021tunable,zhu2022nonlinear}, or parallel resonators~\cite{qu2022underwater,guo2020ultrathin,li2016sound,li2023multifunctional,wang2019extremely,fang2018acoustic,WANG20146828}, with the aim of exceptional acoustic performance subject to the constraint of sample thickness. In these studies involving sound absorption, low-frequency broadband and sub-wavelength evaluation were sometimes used excessively or inappropriately~\cite{ji2020low,guan2023broadband}. 
Luckily, the classic causality theory proposed by Yang et al.~\cite{yang2017optimal} provided a design guideline for broadband absorbers in terms of a thickness limitation of the specimen. Nonetheless, their analysis emphasized mathematical deductions yet without clear physical interpretations, and thus could not explicitly clarify that neither damping nor mass of the system is a determinant factor in sound absorption.

To achieve effective absorption over a wide frequency range, resonators are commonly used in various designs. However, the underlying physical mechanisms of sound absorption are still not fully understood. Consequently, a popular misconception persists, wherein the introduction of extra system mass was erroneously claimed as an enhancement in broadband absorption. In this letter, we present four representative examples to clarify the misconceptions about sound absorption with metamaterials. i) Resonator volume significantly affects sound absorption bandwidth, generally broader in larger resonators; ii) Attaining perfect absorption at very low frequencies is easy but sacrifices absorption bandwidth and lacks acoustic innovation; iii) Coiling structures in resonator cavities is merely one of the strategies to add system mass, similar to neck extension or membrane, and may worsen the system stiffness; iv) The same target area is essential for a fair comparison of sound absorption~\cite{li2016acoustic}. The results reveal that sound absorption performance of a resonator is principally determined by its volume, while the observed low-frequency shift of absorption peak is associated with a trade-off in absorption bandwidth and hence does not represent a real enhancement in sound absorption.

Surface impedance is commonly used for sound absorption calculations which incorporates an imaginary component known as reactance. A zero reactance indicates the resonance condition, which is widely recognized in scientific literature and textbooks.
However, within the frequency range where reactance is non-zero, it lacks a physical interpretation in characterizing absorptive behavior. An available reactance, for example from experiments, can be decomposed into equivalent mass and stiffness terms~\cite{huang2017acoustic,kinsler2000fundamentals}, which provides an innovative approach. This decomposition helps determine the resonance frequency and extract geometric information about the resonator, offering physical insights that reactance alone fails to provide.

Furthermore, many researchers have proposed the presence of negative stiffness or negative bulk modulus~\cite{fang2006ultrasonic,kaina2015negative,kadic20193d,fleury2013extraordinary,cummer2016controlling} as a groundbreaking concept, particularly in the vicinity of resonance frequencies of coupled subsystems.
However, it is crucial to acknowledge that numerical fluctuations unavoidably arise near the frequencies between two resonance peaks, giving rise to the apparent appearance of negative stiffness or negative mass.
Indeed, it is merely a mathematical illusion and does not correspond to any genuine enhancement in sound absorption performance.
 
\textit{Harf-absorption bandwidth.}---A mass-spring-damping system is introduced to characterize the dynamic response of air vibration in sound-absorbing components~\cite{komkin2017sound,lee2019damped,yang2014impact}. The concept of half-absorption bandwidth, once proposed by Maa~\cite{maa1998potential}, is adopted in this study as a new indicator of sound absorption, which is related to the ratio of system damping to mass (see details in supplementary material~\ref{sec:bandwidth}). The optimal damping value, situated between overdamping and underdamping, leads to a unity sound absorption coefficient and is consistently equivalent to the characteristic impedance of the medium, typically air. Hence, half-absorption bandwidth of perfect absorption solely depends on the parameter of system mass. Upon observing this new index, it appears that system stiffness has no impact on the bandwidth. However, when discussing absorption performance, the first step is to specify a resonance frequency, which is determined by the square root of stiffness to mass, as a baseline for comparison. It is crucial to acknowledge that the desired system mass can be easily achieved by adjusting the diameter or length of a resonator neck, while the challenge usually lies in the spacial constraint, i.e., system stiffness, for engineers and researchers. At first glance, a smaller system mass results in a larger half-absorption bandwidth. Nevertheless, once the resonance frequency is fixed, the desired system mass is closely tied to system stiffness. Consequently, it concludes that half-absorption bandwidth is ultimately determined by system stiffness, which is associated with the resonator volume, rather than system mass or damping.

The primary objective of this letter is to explore the physical mechanism and develop an analytical model for sound-absorbing structures, starting from a representative resonator: Helmholtz resonator (HR). It is an effective acoustic attenuation device at low frequencies with the resonance dictated by the equivalent cavity stiffness and equivalent mass~\cite{tang2005helmholtz,bykov2020design,maurel2019enhanced,mercier2017two,wang2019tunable}. 
The classic lumped analysis gives the resonance frequency as $f_\text{res}=(c_0/2\pi)\sqrt{A_{\mathrm{n}}/(Vl_{\mathrm{n}})}$ which is determined by the HR geometry~\cite{selamet2003helmholtz,johansson2001theory}, where $c_0$ is the sound speed of air, $A_{\mathrm{n}}$ the neck cross-sectional area, $V$ the resonator volume, $l_{\mathrm{n}}$ the equivalent neck length. The extension of the neck, variable cross-sectional area, or perforation of neck extension can easily modify the acoustic properties of the HR~\cite{langfeldt2019resonance,rajendran2022design}. The ability to adjust resonance peaks subject to the resonator volume constraint is advantageous in many applications, such as specialized narrow-band noise filtering.

\begin{figure*}[!tbp]
	\centering
	\includegraphics[width=1\textwidth,scale=1]{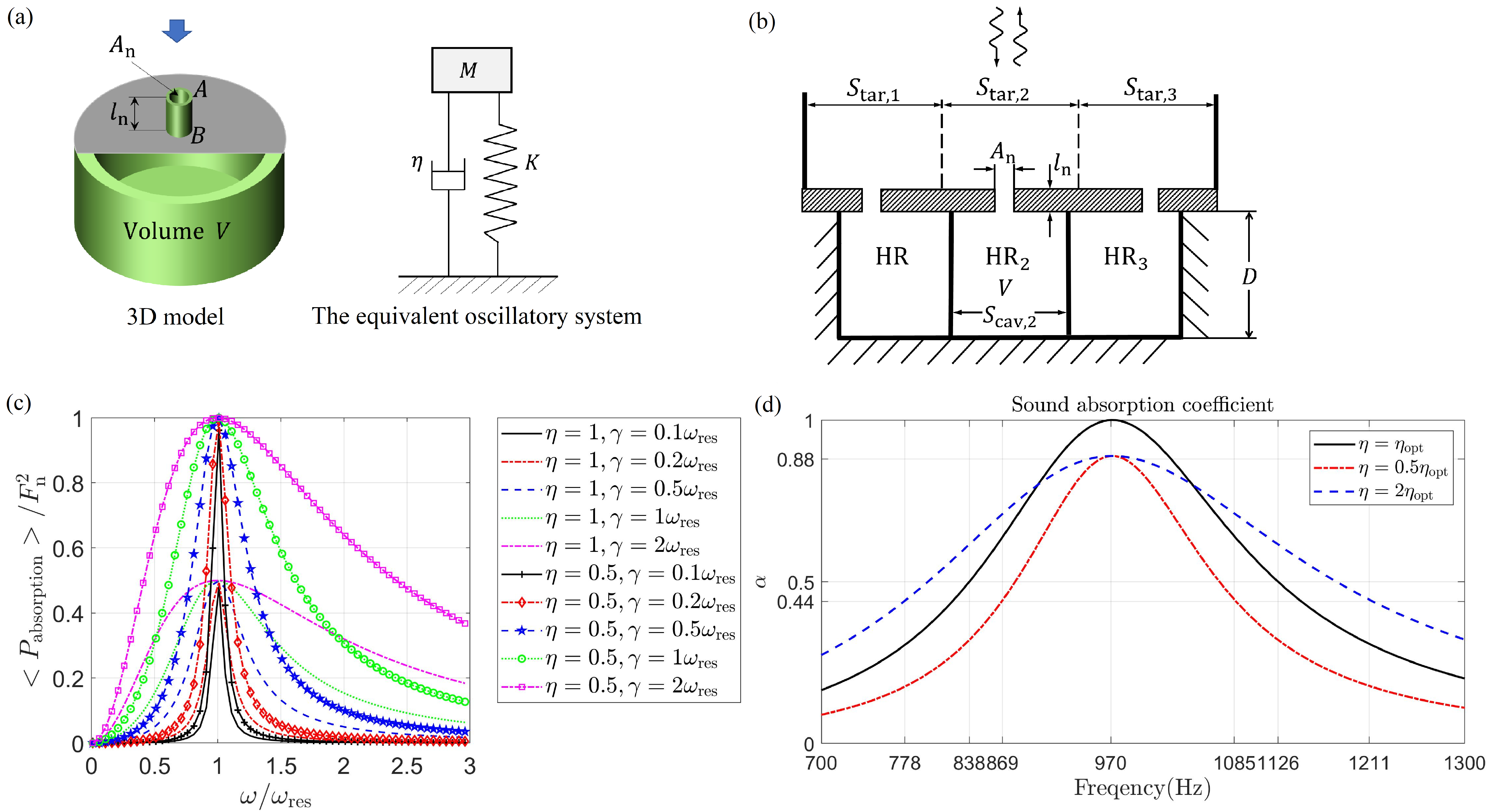}
	\caption{(a) 3D model of an HR and its equivalent oscillatory system, with mass $M$ and damping $\eta$ determined mainly by the neck AB and stiffness $K$ by the cavity volume $V$. (b) 2D schematic of three HRs showing the target area of HR2 labelled as $S_{\mathrm{tar}}$. (c) Frequency spectra of $<P_{\mathrm{absorption}}>$ normalized by $F_{\mathrm{n}}^2$ as in Eq.~\eqref{eq:Pdissipation} for different damping $\eta$ (in units of $\mathrm{kg \,s^{-1}}$). The half-absorption bandwidth $\Delta \omega^{\mathrm{half}}$ of the equivalent oscillatory system is derived in Eq.~\eqref{eq:msd_gamma}. (d) Comparison of absorption peak and bandwidth of a specific HR among three different damping terms $\eta$. The half-absorption bandwidth of the three cases are from $838$ Hz to $1126$ Hz ($\Delta \omega_{\mathrm{HR}}^{\mathrm{half}}=288$~Hz) for optimal damping $\eta_{\mathrm{opt}}$, from $869$ Hz to $1085$ Hz ($\Delta \omega_{\mathrm{HR}}^{\mathrm{half}} = 216$~Hz) for $0.5\eta_{\mathrm{opt}}$, and from $778$ Hz to $1211$ Hz ($\Delta \omega_{\mathrm{HR}}^{\mathrm{half}} = 433$~Hz) for $2\eta_{\mathrm{opt}}$, respectively.}
	\label{Fig1}
\end{figure*}

The three-dimensional (3D) physical model of an HR, as shown in Fig.~\ref{Fig1}(a), can be somehow analogous to a mass-spring-damping oscillatory system. For problems of low-frequency sound absorption, namely the wavelength $\lambda$ greatly larger than the cavity depth $D$, the system mass is dominated by the mass of air within the neck~\cite{kinsler2000fundamentals}, i.e., $M = \rho_0 A_{\mathrm{n}} l_{\mathrm{n}}$, where the mass contribution from the cavity is negligible. The system stiffness $K$ represents the elasticity of air within the cavity, i.e., $K = \rho_0 c_0^2 A_{\mathrm{n}}^2/V$ with $\rho_0$ the density of air.
The resonator can be characterized by an input surface impedance which is the ratio of the external sound pressure at the resonator neck to the velocity averaged across the neck area (see details in supplementary material~\ref{half_absorption-band_width}). The dimensionless input impedance, normalized by $\rho_0c_0$ and related to the neck area $A_\mathrm{n}$, is expressed as in~\cite{komkin2017sound,kinsler2000fundamentals,shen2023absorption}:
\begin{equation} \label{eq:Helm}
	\begin{aligned}
		\bar{Z}_\mathrm{n} =\big[\eta + \mathrm{i} (\omega M - K/\omega)\big]/(\rho_0 c_0 A_{\mathrm{n}}), \\
	\end{aligned}
\end{equation}
where $\omega$ is the angular frequency and the damping of the resonator is given by $\eta = R_{\mathrm{n}}\rho_0 c_0 A_{\mathrm{n}}$ in units of $\mathrm{kg \,s^{-1}}$ (see Eq.~(8) in~\cite{komkin2017sound}), with $R_{\mathrm{n}}=R_{\mathrm{v}}+R_{\mathrm{isot}}$ denotes dimensionless viscous resistance near the neck surface plus dimensionless thermal resistance within the cavity. 
According to Newton's second law, the dynamics of a mass-spring-damping system, such as air vibration in the neck of an HR, can be described by a vibration equation in classical mechanics, $\ddot{x} + \gamma \dot{x} + \omega_{\mathrm{res}}^2 x	= F \cos \omega t$, where $\gamma \equiv \eta/M$ represents the half-absorption bandwidth of the equivalent oscillatory system (see Eq.~\ref{eq:msd_gamma}), with resonance frequency $\omega_{\mathrm{res}} = \sqrt{K/M}$, and $F = F_{\mathrm{n}}/M$ in the form of sound pressure $F_{\mathrm{n}}$ on the resonator neck.
The half-absorption bandwidth is a key index that characterizes the absorption performance of the oscillatory system, as illustrated in Fig.~\ref{Fig1}(c). For more detailed discussions, please refer to supplementary material~\ref{sec:bandwidth}.

The sound absorption coefficient of a specific HR can be obtained as
\begin{equation} \label{eq:alpha_important_prime}
\begin{aligned}
\alpha &= 1 - \left| \dfrac{\bar{Z}_\mathrm{n}/\sigma - 1}{\bar{Z}_\mathrm{n}/\sigma + 1} \right|^2,
\end{aligned}
\end{equation}
where the input impedance $\bar{Z}_\mathrm{n}$ at the neck is divided by the perforation ratio $\sigma = A_{\mathrm{n}}/S_{\mathrm{tar}}$ to represent the total input impedance of the HR.
It is clear from Eq.~\eqref{eq:alpha_important_prime} that sound absorption must be specified on a target area $S_{\mathrm{tar}}$ of the HR, as illustrated in Fig.~\ref{Fig1}(b), which has been validated by the experimental data of Komkin et.al~\cite{komkin2017sound}. 
The optimal damping is not always achievable in practical applications since the damping of porous materials or microholes may deteriorate over long-period usage. The numerical simulations using the commercial software Comsol Multiphysics provide valuable insights into the different damping conditions for a specific HR, highlighting the advantage of an overdamping configuration when optimal damping cannot be achieved, as shown in Fig.~\ref{Fig1}(d). The half-absorption bandwidth of the underdamping ($0.5\eta_{\mathrm{opt}}$) and overdamping ($2\eta_{\mathrm{opt}}$) cases are $216$ Hz and $433$ Hz, respectively, which are 0.75 and 1.5 times the frequency of $288$ Hz in the case of optimal damping ($\eta_{\mathrm{opt}}$). The theoretical validation supporting the conclusion that overdamping is better for sound absorption than underdamping can be found in supplementary material~\ref{half_absorption-band_width}.


\begin{figure*}[!tbp]
	\centering
	\includegraphics[width=0.9\textwidth]{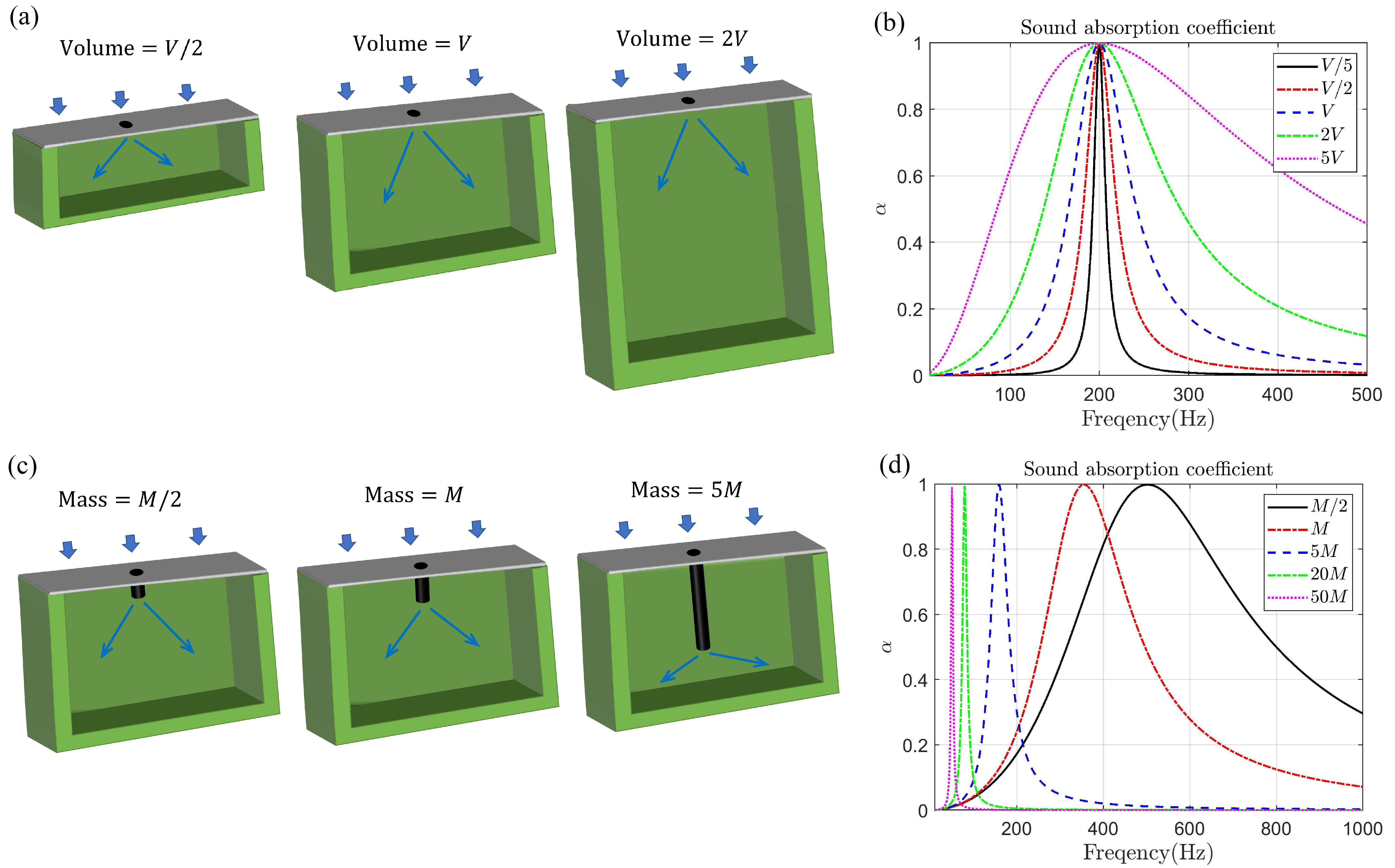}
	\caption{(a) Schematic of three HRs with different volumes. (b) The influence of volume $V$, i.e., system stiffness, on sound absorption coefficient, where mass and damping can be attributed any desirable values. (c) Schematic of three HRs with different neck lengths. (d) Perfect sound absorption can be tuned at any low frequency by properly choosing the system mass $M$ at a fixed volume. This is taken for granted in classic acoustics.}
	\label{Fig4}
\end{figure*}

\textit{Case i: Volume determining absorption (mass disappearing).}---The application of half-absorption bandwidth as an effective indicator of sound absorption is subject to the specification of resonance frequency of the HR. However, sound absorption becomes independent on the mass term once the resonance frequency is fixed, as will be proved subsequently.
The sound absorption coefficient can be expressed as
\begin{equation} \label{eq:alpha_important2}
	\begin{aligned}
		\alpha &=1-\left| \dfrac{1-G_\mathrm{n}\sigma/A_\mathrm{n}}{1+G_\mathrm{n}\sigma/A_\mathrm{n}}\right|^2
	\end{aligned}
\end{equation} 
with the admittance~\cite{shen2023absorption}
\begin{equation} \label{eq:PPTG0}
	\begin{aligned}
		G_\mathrm{n} = \dfrac{A_\mathrm{n} }{\bar{Z_\mathrm{n}}}
		&= \dfrac{A_\mathrm{n} }{\dfrac{l_\mathrm{n} R_0}{\rho_0 c_0} + \dfrac{\mathrm{i} \omega l_\mathrm{n} }{c_0} + \dfrac{A_\mathrm{n} c_0}{\mathrm{i} \omega V} }\\
		&= \dfrac{\omega V/c_0 }{\dfrac{R_0\omega}{\rho_0\omega_{\mathrm{res}}^2} + \mathrm{i} \left(\dfrac{\omega^2}{\omega_{\mathrm{res}}^2}- 1 \right)}\\
		&= \dfrac{\mathrm{i}k V}{1-\left(\dfrac{f}{f_{\mathrm{res}}}\right)^2 + \mathrm{i}\left(\dfrac{f}{f_{\mathrm{res}}}\right)  \left(\dfrac{R_0}{2 \pi f_{\mathrm{res}} \rho_0}\right) },
	\end{aligned}
\end{equation}
where the equivalent characteristic flow resistivity $R_0=\eta/(A_\mathrm{n}l_\mathrm{n})$, i.e., the ratio of system damping to neck volume. It is always possible to find an optimal damping value, determined by the flow resistivity $R_0$ of the neck, to achieve perfect absorption. 
Although the resonator neck and cavity volume together determine the resonance frequency $f_{\mathrm{res}}$, the resonator's absorption performance depends solely on the volume $V$ at the fixed $f_{\mathrm{res}}$. The mass term explicitly disappears in Eq.~\eqref{eq:PPTG0}, which is a significant finding of this letter.

In Fig.~\ref{Fig4}(a), three HRs featuring rectangular cavities of different volumes are presented. The acoustic performance of an HR is nearly insensitive to the geometry of a cavity with uniform sound pressure (i.e., acoustically compact). As depicted in Fig.~\ref{Fig4}(b), the resonance frequencies of the three HRs are all set to $200$ Hz, and the system damping can be easily adjusted through the flow resistivity $R_0$ to achieve a unity absorption coefficient at the resonance frequency.
Therefore once the resonance frequency $f_{\mathrm{res}}$ of an HR is fixed, the spectral shape of absorption coefficient, and hence the half-absorption bandwidth, is determined by the volume $V$, as shown in Fig.~\ref{Fig4}(b). Despite numerous previous studies focusing on mass addition or mass modification~\cite{mei2012dark,long2019broadband,liu2000locally,hong2006sound} to adjust the absorption peak and claiming broadband absorption or performance enhancement, the `mass disappearing' phenomenon, as evident in Eq.~\eqref{eq:PPTG0}, reaffirms that volume or system stiffness is the only determinant factor in sound absorption performance.

\textit{Case ii: Trade-off of low frequency and bandwidth.}---Given that half-absorption bandwidth is closely related to resonance frequency, it is crucial to further investigate whether shifting the resonance peak to low frequencies can be deemed as an innovation with practical meaning~\cite{mei2012dark,assouar2018acoustic}. The resonance frequency is obtained by 
\begin{equation} \label{eq:fres2}
	\begin{aligned}
		f_\mathrm{res}  = \dfrac{1}{2\pi}\sqrt{\dfrac{K}{M}}= \dfrac{c_0}{2\pi}\sqrt{\dfrac{A_\mathrm{n}}{Vl_\mathrm{n}}}.
	\end{aligned}
\end{equation}
Figure~\ref{Fig4}(c) showcases three HRs with varying neck length $l_\mathrm{n}$, which determines the equivalent mass $M$ of oscillating resonators. 
Perfect sound absorption at lower frequencies becomes feasible by increasing the system mass, as depicted in Fig.~\ref{Fig4}(d), in which a near unity coefficient of $\alpha = 1$ is obtained by adjusting the damping term $R_0$.
It concludes that absorption peak at arbitrarily low frequencies is taken for granted in classic acoustics, as seen in Eq.~\eqref{eq:fres2} by tuning the neck length. It is evident in Fig.~\ref{Fig4}(d) that when the total volume is kept constant, achieving perfect absorption at low frequencies narrows the half-absorption bandwidth (see details in Eq.~\ref{eq:gamma_HR}). Therefore, discussing `sub-wavelength' performance at discrete frequencies may not be meaningful due to the trade-off between the resonance peak shifting to low frequencies and the resultant sacrifice in absorption bandwidth.

\begin{figure*}[!htbp]
	\centering
	\includegraphics[width=0.9\textwidth]{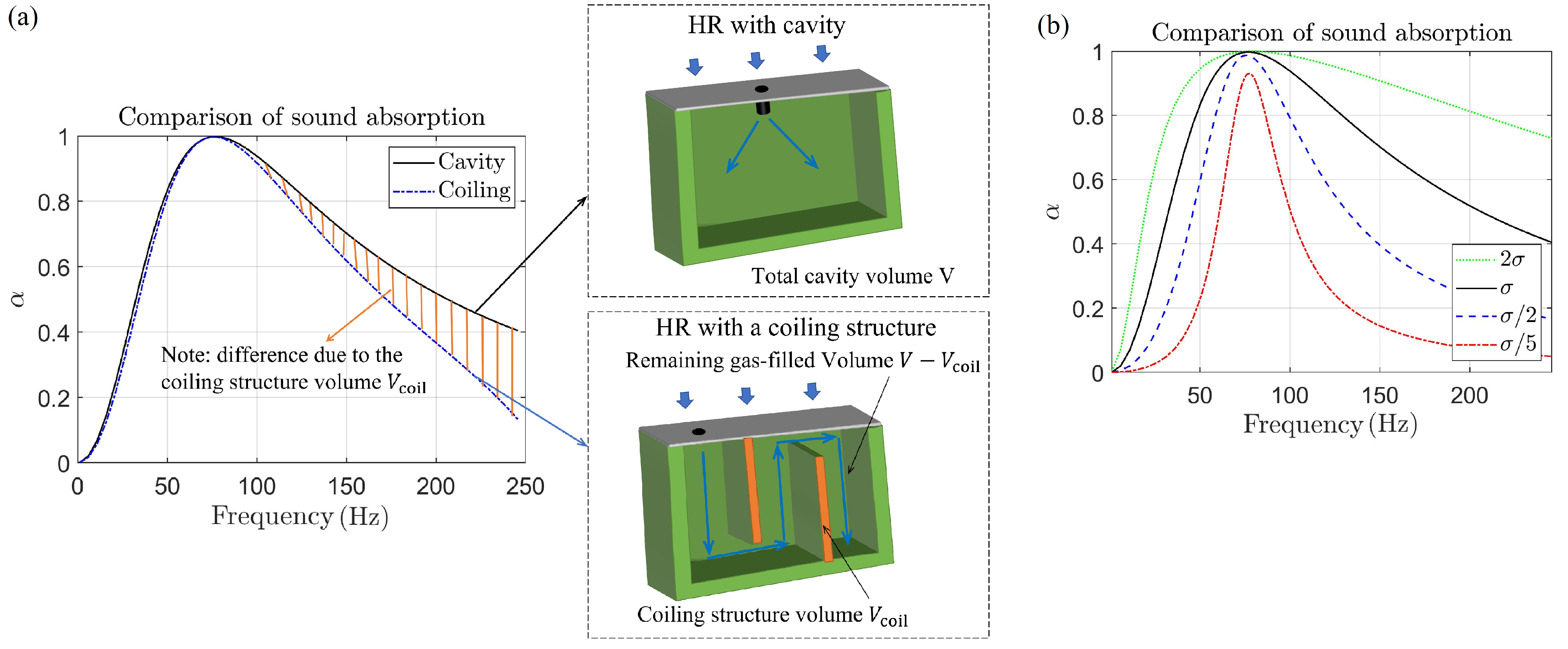}
	\caption{(a) Comparison of the optimal absorption between two HRs backed with a cavity only and a coiling structure, constrained with a given resonance and the same total volume. (b) Comparison of sound absorption when the input surface impedance is specified on different target areas, i.e., different perforation ratios $\sigma$,  with the same cavity volume.}
	\label{Fig5}
\end{figure*}

\textit{Case iii: `Perfect absorption' by coiling structures.}---Coiling structures have also received great research attentions~\cite{li2016acoustic,gao2021hybrid,wu2019low,LiuChenkai2021,liang2012extreme,long2019broadband,long2019broadband,landi2018acoustic,cheng2015ultra} as an innovative way to modify the system mass of an HR.
The presence of a coiling structure elongates the propagation path of sound waves within the cavity, i.e., $l_{\mathrm{coil}} \gg D$ with $l_{\mathrm{coil}}$ the length of the coiling structure, and hence a larger equivalent cavity depth makes an extra contribution to the system mass. Advanced mathematical techniques, such as Taylor series expansion, can be employed to model and interpret the physical meaning of coiling structures. The reactance of a coiling structure can be expanded as~\cite{huang2017acoustic,shen2023absorption}
\begin{equation} \label{eq:Taylor}
	\begin{aligned}
 \mathrm{Im}(Z_{\mathrm{coil}}) &= - \mathrm{i} \rho_0 c_0 \cot (k l_{\mathrm{coil}})\\
			&\approx  \mathrm{i}\omega \left( \dfrac{\rho_0 l_{\mathrm{coil}}}{3} -\dfrac{\rho_0 c^2_0/l_{\mathrm{coil}}}{\omega^2 } \right),
	\end{aligned}
\end{equation}
where higher-order terms in the expansion have been ignored for compact structures (i.e., $\omega l_{\mathrm{coil}}/c_0 \ll 1$).
The first term $\rho_0l_{\mathrm{coil}}/3$ in Eq.~\eqref{eq:Taylor} indicates that approximately one-third of the mass of air in the coiling structure equivalently participates in the system resonance.

The HR under study consists of both solid coiling plates highlighted in orange and an air-filled cavity, as shown in Fig.~\ref{Fig5}(a). However, the low-frequency `perfect absorption' attributed to the coiling structure can inherently be alternatively achieved by extending the neck length, which similarly serves to increase the overall system mass. 
The total volume of the HR with coiling structures, denoted as $V$, can be divided into the volume $V-V_{\mathrm{coil}}$ occupied by air and the volume $V_{\mathrm{coil}}$ of the solid coiling plates which is a small fraction of the overall space.
This difference in air-filled volume between the HRs with and without the coiling structure slightly increases the stiffness of the equivalent oscillatory system for the coiling-structure HR, denoted as $K = \rho_0 c_0^2 A_{\mathrm{n}} /(V-V_{\mathrm{coil}})$.
As a consequence, the coiling-structure HR tends to exhibit a comparatively inferior overall performance compared to the HR with an extended neck in achieving optimal sound absorption at a fixed resonance frequency, particularly at frequencies higher than $f_{\mathrm{res}}$, as illustrated in Fig.~\ref{Fig5}(a).

\textit{Case iv: Fixing target area.}---The sound absorption and input impedance of an HR must be defined for a specific target area, as can be seen in Eq.~\eqref{eq:alpha_important_prime}, which is related to the perforation ratio as $\sigma=A_{\mathrm{n}}/S_{\mathrm{tar}}$.
Consequently, the sound absorption coefficient varies when the HR is placed on a different target area while maintaining the same resonant cavity, as depicted in Fig.~\ref{Fig5}(b). More details regarding the target area can be found in supplementary material~\ref{half_absorption-band_width}.
It is important to note that choosing an improper target area $S_{\mathrm{tar}}$ can lead to a misleading illusion of enhanced sound absorption (for example~\cite{li2016acoustic}).

\textit{Discussion.}---Another significant and innovative byproduct of the mass-spring-damping model is to partition reactance. 
By representing reactance as equivalent mass and stiffness terms in the form of Eq.~\eqref{eq:Helm}, it enables a systematic estimation of neck size and cavity volume. Parameter fitting of reactance data across a range of frequencies helps extract effective stiffness and damping parameters for an HR, thereby relating reactance to resonator dynamics in both resonant and non-resonant regimes. Supplementary material~\ref{sec:Quantifying} provides detailed information on applying least-mean-square fits to decompose the mass-spring-damping model through reactance~\cite{huang2017acoustic}. In summary, partitioning reactance offers physical insights beyond the reactance alone and allows the prediction of the resonator's geometric properties based on its acoustic characteristics.

It is important to note that fitting over a frequency band between two resonance peaks might generate mathematically negative equivalent stiffness or bulk modulus (see examples~\cite{fang2006ultrasonic,kaina2015negative,kadic20193d,fleury2013extraordinary,cummer2016controlling}). However, this does not indicate the presence of negative stiffness in real world. Bobrovnitskii et al.~\cite{bobrovnitskii2014effective} concluded that assessing the energy of a sound wave was not solely dependent on the equivalent mass and stiffness but also on their frequency derivatives, specifically $\mathrm{d} M/\mathrm{d} \omega$ and $\mathrm{d} K/\mathrm{d} \omega$. 
Moreover, a comparison of simulation results shows a total absorption of two HRs in parallel less than the sum of the individual HR's absorption, which indicates that the overall system does not exhibit true negative stiffness behavior despite numerical singularities (see details in supplementary material~\ref{sec:Quantifying}). In fact, passive structures relying solely on interactions or geometry modifications, without involving other physical fields, cannot give rise to real negative stiffness.

The mass-spring-damping model also provides insights into the mechanism of inter-resonator coupling between resonators assembled in parallel configurations~\cite{shen2021acoustic} (see details in supplementary material~\ref{sec:coupling}), which was still unclear despite numerous previous studies~\cite{yang2017optimal,qu2022underwater,guo2020ultrathin,li2016sound,zhang2021broadband,li2023multifunctional,wang2019extremely,fang2018acoustic}. The problem of inter-resonator coupling leads to a redistribution of the overall mass and damping characteristics of the system~\cite{auregan2019parallel}. 
One of the fundamental techniques used in this study involves extracting near-field sound pressure and intensity maps by subtracting the far-field quantities from the total sound field. These maps illustrate the influence of near-field coupling on the distribution of sound pressure and internal acoustic intensity flow in the near-field region. The inter-resonator coupling is equivalent to a virtual resonator formed between two inter-connected resonators, where the one with a lower resonance frequency acts as a mass and the other as a spring. This inter-connected resonance occupies a portion of the vibrating air but does not contribute to the far-field, resulting in a reduction of mass. The reduced mass leads to a wider half-absorption bandwidth compared with each individual resonator, which effectively explains the observed absorption performance in the valleys between resonance peaks~\cite{yang2017optimal}.

\textit{Conclusion.}---In summary, this letter presents an oscillatory-diffusive representation of sound absorption structures using a surface impedance formulation. It also derives expressions for equivalent mass, stiffness, and damping specific to sound absorption models. Notably, this study highlights the misleading concept of `sub-wavelength' in evaluating sound absorption and introduces the half-absorption bandwidth as a new evaluation criterion. Furthermore, in practical engineering applications, the damping term often inevitably differs from its optimal value. In such cases, overdamping enhances overall sound absorption performance in contrast to underdamping.

The decomposition of the imaginary part of surface impedance into mass and stiffness components allows for a state-of-the-art interpretation of the physical mechanism of the inter-resonator coupling between resonators in parallel.
It is unfair to compare the performance of assembly non-uniform resonators in a narrow stopband with the apparently low absorption of a broadband device operating at frequencies far away from its own resonance. Attempts to enhance sound absorption performance by solely adjusting system mass through geometry and structural designs ultimately result in a trade-off, in that low-frequency improvements lead to sacrifices in bandwidth which is primarily determined by the resonator volume, i.e., system stiffness. It can be argued that such types of structures are better described as bandstop filters rather than sound absorbers for broadband noise. We conclude that many newly proposed sound-absorbing structures labeled as $\mathit{metamaterial}$ essentially lack technical breakthroughs, and the descriptor $sub$-$wavelength$, as a common phenomenon in acoustics, is merely an overestimated innovation. This insight opens up new possibilities for the design of sound-absorbing structures, as it shifts the goal from adding mass to reducing stiffness in order to achieve desired low-frequency broadband absorption. However, when fitting a mathematical model over the frequency band between two resonance peaks, the claimed `negative' stiffness or bulk modulus does not imply that the material response is physical since there is no real performance improvement.

C. Shen expresses gratitude to Southern University of Science and Technology (SUSTech) for the full scholarship to support his research under the HKU-SUSTech joint Ph.D. program.

\appendix
\setcounter{table}{0} 
\setcounter{figure}{0}
\renewcommand{\thetable}{A\arabic{table}}
\renewcommand{\thefigure}{A\arabic{figure}}

\section{Theoretical formulation}
\subsection{Half-absorption bandwidth of a mass-spring-damping model}
\label{sec:bandwidth}
According to Newton's second law, the mass-spring-damping system of an HR is governed by 
$F_{\mathrm{spring}}+ F_{\mathrm{damping}} +  F_{\mathrm{driving}} = M a$, yielding
\begin{equation} \label{eq:MKD}
	\begin{aligned}
		M \ddot{x} + \eta \dot{x} + K x	= F_{\mathrm{d}} \cos \omega t.
	\end{aligned}
\end{equation}
By assigning the sound pressure $F_{\mathrm{n}}$ at the neck of the HR as the driving force $F_{\mathrm{d}}$, Eq.~\eqref{eq:MKD} can be rewritten as
\begin{equation} \label{eq:MKD2}
	\begin{aligned}
		\ddot{x} + \gamma \dot{x} + \omega_{\mathrm{res}}^2 x	= F \cos \omega t,
	\end{aligned}
\end{equation}
where 
\begin{equation} \label{eq:msd_gamma}
	\begin{aligned}
		\gamma \equiv \eta/M,	
	\end{aligned}
\end{equation}
the system resonance frequency $\omega_{\mathrm{res}} = \sqrt{K/M}$, and $F = F_{\mathrm{n}}/M$.
These equations provide a mathematical representation of the dynamic behavior of an HR in response to the applied driving force.

Note that the solution $x(t)$ of a vibratory system can be shown in multiple equivalent expressions~\cite{morin2008introduction}
\begin{equation} \label{eq:xt}
	\begin{aligned}
		x(t) &= A \cos(\omega t + \phi)\\
		&= A \sin(\omega t + \phi^{\prime})\\
		&= B_{\mathrm{c}} \cos \omega t  + B_{\mathrm{s}} \sin \omega t \\
		&= C \mathrm{e}^{\mathrm{i} \omega t } + C^{*} \mathrm{e}^{-\mathrm{i} \omega t}\\
		&= \mathrm{Re} (D \mathrm{e}^{\mathrm{i} \omega t }).
	\end{aligned}
\end{equation}
One form in Eq.~\eqref{eq:xt} might work better than the others depending on the situation, such as physical interpretation or mathematical deduction.\par
In an HR system, the pressure in the neck acts as the driving force that feeds sound energy into the resonator. Simultaneously, the damping force, represented by $-\eta \dot{x}$ where $\eta$ is the damping coefficient and $\dot{x}$ the velocity, is always opposite to the motion of the medium. There is an exception at the resonance frequency when it always absorbs energy. Considering a steady-state solution, the net power from the incident pressure, averaged over a period, is equal to the power dissipated by the damping force.

Recalling the equation of motion, $x(t) = A \cos(\omega t + \phi)$, we can calculate the power dissipated by the damping force as 
\begin{equation} \label{eq:Pdamping}
	\begin{aligned}
		P_{\mathrm{damping}} = (-\eta \dot{x}) \dot{x} 
		= -\eta (\omega A)^2 \sin^2(\omega t + \phi).
	\end{aligned}
\end{equation}
Integrating this equation over a complete period gives
\begin{equation} \label{eq:PdampingAverage}
	\begin{aligned}
		<P_{\mathrm{damping}}>= - \frac{1}{2}\eta (\omega A)^2.
	\end{aligned}
\end{equation} 
Likewise, we can calculate the power supplied by the incident pressure as
\begin{equation} \label{eq:Pdriving}
	\begin{aligned}
		P_{\mathrm{driving}} &= (F_{\mathrm{n}} \cos \omega t ) \dot{x} \\
			&= - F_{\mathrm{n}} \omega A \cos \omega t (\sin \omega t  \cos \phi + \cos \omega t  \sin \phi).
	\end{aligned}
\end{equation}
Equations~\eqref{eq:Pdamping} and ~\eqref{eq:Pdriving} do not cancel each other all the time but sum up to zero on average. Repeating the same integration for the driving power over one period, only the second term in the bracket of Eq.~\eqref{eq:Pdriving} remains with $<\cos^2 (\omega t)> = 1/2$. Thus, the integrated form of Eq.~\eqref{eq:Pdriving} is reduced to
\begin{equation} \label{eq:PdrivingAverage}
	\begin{aligned}
		<P_{\mathrm{driving}}> = -\dfrac{1}{2} F_{\mathrm{n}} \omega A \sin \phi .
	\end{aligned}
\end{equation} 

To obtain the term $\sin \phi$, assuming the frequency of motion is equal to the driving frequency and substituting $x(t)$ into Eq.~\eqref{eq:MKD2} yield
\begin{equation} \label{eq:xt2}
	\begin{aligned}
		\omega^2 A \cos(\omega t + \phi +\pi) + \gamma \omega A \cos(\omega t + \phi +\pi/2) \\
		 +\omega_{\mathrm{res}}^2 A \cos(\omega t + \phi ) = F \cos \omega t,
	\end{aligned}
\end{equation}
which indicates that acceleration is $90^{\circ}$ ahead of velocity and $180^{\circ}$ ahead of displacement in phase.
The phase $\phi$ in Eq.~\eqref{eq:xt2} can be obtained as $\tan \phi = -\gamma \omega/(\omega_{\mathrm{res}}^2 - \omega^2)$.
Taking a mean square of Eq.~\eqref{eq:xt2} gives
\begin{equation} \label{eq:A}
	\begin{aligned}
		((\omega_{\mathrm{res}}^2 - \omega^2)A)^2 + (\gamma \omega A)^2 = F^2,
	\end{aligned}
\end{equation}
and so the amplitude is obtained as
 \begin{equation*} \label{eq:A2}
  A = F/(\sqrt{(\omega_{\mathrm{res}}^2 - \omega^2)^2 + (\gamma \omega)^2}).
 \end{equation*}
The term $\sin \phi$ in Eq.~\eqref{eq:PdrivingAverage} is then derived as $\sin \phi = -\gamma \omega A/F = -\gamma M \omega A/F_{\mathrm{n}}$.
Substituting it into Eq.~\eqref{eq:PdrivingAverage}, one obtains
\begin{equation} \label{eq:PdrivingAverage2}
	\begin{aligned}
		<P_{\mathrm{driving}}> = -\dfrac{1}{2} F_{\mathrm{n}} \omega A (\dfrac{-\gamma M \omega A}{F_{\mathrm{n}}}) 
		= \dfrac{1}{2}\eta (\omega A)^2.
	\end{aligned}
\end{equation}

To explain sound absorption from the perspective of physical phenomena using Eqs.~\eqref{eq:PdampingAverage} and ~\eqref{eq:PdrivingAverage2}, the sound energy absorbed by the HR system during each integration period is 
\begin{equation} \label{eq:Pdissipation}
	\begin{aligned}
		<P_{\mathrm{absorption}}> &= <P_{\mathrm{driving}}> = - <P_{\mathrm{damping}}> \\
		&= \dfrac{\eta \omega^2}{2}  \dfrac{(F_{\mathrm{n}}/M)^2}{(\omega_{\mathrm{res}}^2 - \omega^2)^2 + (\gamma \omega)^2} \\
		&= \dfrac{F_{\mathrm{n}}^2}{2 \eta}
		\fbox{$\dfrac{\gamma^2 \omega^2}{(\omega_{\mathrm{res}}^2 - \omega^2)^2 + (\gamma \omega)^2} $} \\
		& \equiv \dfrac{F_{\mathrm{n}}^2}{2 \eta}  \fbox{$f(\omega)$},
	\end{aligned}
\end{equation}
where $\gamma$ is defined in Eq.~\eqref{eq:msd_gamma}, in units of $\mathrm{s^{-1}}$.

There are several interesting characteristics of the function $f(\omega)$ in Eq.~\eqref{eq:Pdissipation}, as depicted in Fig.~\ref{Fig1}(c). This function reaches its maximum of 1 at the resonance frequency $\omega_{\mathrm{res}}$. A critical condition occurs when the absorption power is attenuated by 3 dB, i.e., $f(\omega) = 0.5$. In this case, the two terms in the denominator of $f(\omega)$ must be equal, leading to
\begin{equation} \label{eq:omega0}
	\begin{aligned}
		\omega_{\mathrm{res}}^2 - \omega^2 =  \pm \gamma \omega,
	\end{aligned}
\end{equation}
and hence the critical frequencies for a halved absorption power can be determined from Eq.~\eqref{eq:omega0}.

Provided that the resonance frequency $\omega_{\mathrm{res}}$ of the oscillatory system is already known, the index of half-absorption bandwidth as proposed in this letter can be calculated. Setting the two critical frequencies for half absorption in Eq.~\eqref{eq:omega0} as $\omega_1$ and $\omega_2$, i.e.,
\begin{equation*} \label{eq:omega12}
	\begin{aligned}
		\omega_{\mathrm{res}}^2 - \omega_1^2 &=  \gamma \omega_1, \\
		\omega_{\mathrm{res}}^2 - \omega_2^2 &=  -\gamma \omega_2,
	\end{aligned}
\end{equation*}
and
\begin{equation*} \label{eq:omega_difference}
	\begin{aligned}
		\omega_2^2 -\omega_1^2 = \gamma (\omega_2+\omega_1),
	\end{aligned}
\end{equation*}
the half-absorption bandwidth $\Delta \omega^{\mathrm{half}}$ is obtained as
\begin{equation} \label{eq:omega}
	\begin{aligned}
		\Delta \omega^{\mathrm{half}} \equiv \omega_2 -\omega_1 = \gamma.
	\end{aligned}
\end{equation}
It is interesting to observe that the quantity $\Delta \omega^{\mathrm{half}}$ coincides precisely with the parameter $\gamma$, which is defined in Eq.~\eqref{eq:msd_gamma} as the ratio of system damping $\eta$ to system mass $M$. This remarkable relation holds true for any arbitrary values of $\gamma$.

\subsection{Half-absorption bandwidth of an HR}
\label{half_absorption-band_width}
%
HR, as shown in Fig.~\ref{Fig1}(a), is one of the most useful devices in sound absorption. It consists of a cavity with volume $V$ and a short neck $\mathrm{AB}$ which has a cross-sectional area $A_n$, a radius $r_0$, and a physical neck length $l_{\mathrm{n}}^{\prime}$; the effective neck length $l_n=l_{\mathrm{n}}^{\prime}+l_{\mathrm{n,virtual}}$, where $l_{\mathrm{n,virtual}}$ is the end correction length of approximately $0.85r_0$ or $0.60r_0$ depending on the presence or absence of a flange. 
At low frequencies, the cavity can be regarded as a compact air spring with uniform pressure $p_{\mathrm{B}}$ and density variation. Mass conservation within the HR can be expressed as
\begin{equation} \label{eq:PPTmass}
	\begin{aligned}
		\rho_0 u A_{\mathrm{n}} = V \dfrac{\partial \rho}{\partial t} = \dfrac{\mathrm{i} V\omega p_\mathrm{B}}{c_0^2},
	\end{aligned}
\end{equation}
where $p_\mathrm{B} = \rho_0 c_0^2 u A_{\mathrm{n}} /(\mathrm{i} \omega V)$ and $u$ denotes the acoustic particle velocity.

Assume that air particles vibrate with uniform velocity $u$ along the neck, namely, a damping material with flow resistivity $R_0 \equiv \eta/(A_{\mathrm{n}}l_{\mathrm{n}})$ which has been defined in Eq.~\eqref{eq:PPTG0} and is related with particle velocity in another form $R_0=-u^{-1}\mathrm{d}p/\mathrm{d}x$ in units of $\mathrm{kg} \,\mathrm{m}^{-3} \,\mathrm{s}^{-1}$. Thus, momentum conservation in the neck can be written as follows:
\begin{equation} \label{eq:PPTMomentum}
	\begin{aligned}
		p_\mathrm{A} - p_\mathrm{B} - u l_{\mathrm{n}} R_0 &= \rho_0 A_{\mathrm{n}} l_{\mathrm{n}} \dfrac{\partial u}{\partial t}= \mathrm{i} \rho_0 A_{\mathrm{n}} l_{\mathrm{n}}  \omega u. 
	\end{aligned}
\end{equation}
The dimensionless input impedance of an HR, normalized by $\rho_0c_0$, is dependent on the neck area $A_\mathrm{n}$ and can be expressed as
\begin{equation} \label{eq:PPTZ}
	\begin{aligned}
		\bar{Z}_\mathrm{n} = \dfrac{p_\mathrm{A}}{\rho_0 c_0 u}
		&= \dfrac{\eta}{\rho_0 c_0A_\mathrm{n}}+ \mathrm{i} k l_\mathrm{n} + \dfrac{A_\mathrm{n}}{\mathrm{i} k  V},
	\end{aligned}
\end{equation}
which is consistent with Eq.~\eqref{eq:Helm} with the wavenumber $k=\omega/c_0$. The resonance occurs when the imaginary part of the input impedance is zero, that is, $\mathrm{Im}(Z_\mathrm{n})=0$; thus, $\omega_{\mathrm{res}} = 2 \pi f_{\mathrm{res}}= c_0 \sqrt{A_\mathrm{n} /(V l_\mathrm{n}) }$.

The admittance of the HR, which is dimensional in units of $\mathrm{m}^{2}$ on the neck, can be defined as $G_\mathrm{n} = A_\mathrm{n} /\bar{Z}_\mathrm{n}$.
The sound absorption coefficient should be defined on a specified target area $S_{\mathrm{tar}}$, as shown in Fig.~\ref{Fig1}(b), i.e.,
\begin{equation} \label{eq:alpha_important}
	\begin{aligned}
		\alpha &=1-\left| \dfrac{\bar{Z}_\mathrm{n}/\sigma-1}{\bar{Z}_\mathrm{n}/\sigma+1}\right|^2,
	\end{aligned}
\end{equation}
where the dimensionless impedance $\bar{Z}_\mathrm{n}$ of the HR in Eq.~\eqref{eq:PPTZ} is related to this target area through the perforation ratio $\sigma = A_{\mathrm{n}}/S_{\mathrm{tar}}$.

Equation~\eqref{eq:alpha_important} can be rewritten in terms of the admittance $G_\mathrm{n}$ given in Eq.~\eqref{eq:PPTG0} as
\begin{equation} \label{eq:alpha_Gn1}
	\begin{aligned}
		\alpha 	&=1-\left| \dfrac{S_\mathrm{tar}-G_\mathrm{n}}{S_\mathrm{tar}+G_\mathrm{n}}\right|^2.
	\end{aligned}
\end{equation}
Replacing $G_\mathrm{n}$ with an alternative simplified form $G_\mathrm{n}=\mathrm{i}a/(b+\mathrm{i}c)$, where $a\equiv \omega_{\mathrm{res}}^2 \omega V/c_0$, $b\equiv \omega_{\mathrm{res}}^2-\omega^2$, and $c\equiv \omega \eta/M$, Eq.~\eqref{eq:alpha_Gn1} becomes
\begin{equation} \label{eq:alpha_Gn2}
	\begin{aligned}
		\alpha &=1-\left| \dfrac{S_\mathrm{tar}-\dfrac{\mathrm{ia}}{b+\mathrm{i}c}}{S_\mathrm{tar}+\dfrac{\mathrm{ia}}{b+\mathrm{i}c}}\right|^2
		=\dfrac{4S_\mathrm{tar}ca}{(S_\mathrm{tar}b)^2+(S_\mathrm{tar}c+a)^2}.
	\end{aligned}
\end{equation}
Substituting the expressions of $a$, $b$ and $c$ back into the above equation yields
\begin{equation} \label{eq:alpha_Gn3}
	\begin{aligned}
		\alpha 	&=\dfrac{4S_\mathrm{tar} \omega_{\mathrm{res}}^2 \omega^2 (\eta/M) V/c_0}{S_\mathrm{tar}^2(\omega_{\mathrm{res}}^2-\omega^2)^2+\omega^2(S_\mathrm{tar} \eta/M+ \omega_{\mathrm{res}}^2 V/c_0)^2}.
	\end{aligned}
\end{equation}
For comparison of sound absorption, the resonance frequency $\omega_{\mathrm{res}}$ must be first specified. When $\omega = \omega_{\mathrm{res}}$, the sound absorption coefficient reaches its peak,
\begin{equation} \label{eq:alpha_maximum}
	\begin{aligned}
		\alpha_{\mathrm{p}} 
		&=\dfrac{4S_\mathrm{tar} \omega_{\mathrm{res}}^2 (\eta/M ) V/c_0}{(S_\mathrm{tar}\eta/M + \omega_{\mathrm{res}}^2 V/c_0)^2}.
	\end{aligned}
\end{equation}
The maximum absorption peak of $\alpha_{\mathrm{p}}=1$ can be obtained by taking the optimal damping of the HR,
\begin{equation} \label{eq:eta_opt}
    \begin{aligned}
        \eta_{\mathrm{opt}}\equiv \omega_{\mathrm{res}}^2 MV/(c_0S_\mathrm{tar}) = \eta \sigma/R_{\mathrm{n}},
    \end{aligned}
\end{equation}
which implies that the dimensionless damping term $R_{\mathrm{n}}$ is exactly equal to the perforation ratio $\sigma$ to achieve the optimal damping. 

Therefore, the half-absorption bandwidth $\Delta \omega_{\mathrm{HR}}^{\mathrm{half}}$ of the HR can be evaluated by taking $\alpha=0.5\alpha_{\mathrm{p}}$ in Eq.~\eqref{eq:alpha_Gn3}. This leads to the following equation similar to Eq.~\eqref{eq:omega0},
\begin{equation} \label{eq:omega0_HR}
	\begin{aligned}
		\omega_{\mathrm{res}}^2 - \omega^2 = \pm \gamma_{\mathrm{HR}} \omega,
	\end{aligned}
\end{equation}
where the term
\begin{equation} \label{eq:gamma_HR}
	\begin{aligned}		 
         \gamma_{\mathrm{HR}}\equiv \dfrac{\eta}{M}+ \dfrac{V}{c_0S_\mathrm{tar}} \omega_{\mathrm{res}}^2=\dfrac{\eta}{M}+\dfrac{\eta_{\mathrm{opt}}}{M} 
	\end{aligned}
\end{equation}
is in units of $\mathrm{s^{-1}}$. Following the similar procedure from Eq.~\eqref{eq:omega0} to Eq.~\eqref{eq:omega}, the half-absorption bandwidth $\Delta \omega_{\mathrm{HR}}^{\mathrm{half}}$ of the HR is obtained as 
\begin{equation} \label{eq:omega_HR}
	\begin{aligned}
		\Delta \omega_{\mathrm{HR}}^{\mathrm{half}} \equiv \omega_2 -\omega_1 = \gamma_{\mathrm{HR}}.
	\end{aligned}
\end{equation} 
It is important to note that $\eta/M$ represents the half-absorption bandwidth of the equivalent oscillatory system as has been derived in supplementary material~\ref{sec:bandwidth}.  
When the system damping $\eta$ is tuned to the optimal value through $R_{\mathrm{n}} = \sigma$, the corresponding half-absorption bandwidth in Eq.~\eqref{eq:gamma_HR} becomes $\gamma_{\mathrm{HR,opt}} = 2\eta_{\mathrm{opt}}/M$. 
More importantly, when the resonance frequency $\omega_{\mathrm{res}}$ and absorption peak $\alpha_{\mathrm{p}}$ are specified, sound absorption performance is dominated by the bandwidth $\Delta \omega_{\mathrm{HR}}^{\mathrm{half}}$, which is jointly determined by the volume $V$ of the resonant cavity and the target area $S_\mathrm{tar}$, as shown in Eq.~\eqref{eq:gamma_HR}. The influence of target area on the optimal damping and half-absorption bandwidth is evident, as from Eqs.~\eqref{eq:eta_opt} and \eqref{eq:gamma_HR}, but is usually ignored in previous studies.

There is a compromise between the sound absorption peak in Eq.~\eqref{eq:alpha_maximum} and half-absorption bandwidth when varying the damping from its optimal value, as shown in Fig.~\ref{Fig1}(d). 
Let's examine the two scenarios below:\\
i) When the damping of the HR is set to $2\eta_{\mathrm{opt}}$, the absorption peak drops to $8/9$, while the half-absorption bandwidth increases to 1.5$\gamma_{\mathrm{HR,opt}}$ .\\
ii) When the damping of the HR is set to $0.5\eta_{\mathrm{opt}}$, the absorption peak also drops to $8/9$, while the half-absorption bandwidth decreases to 0.75$\gamma_{\mathrm{HR,opt}}$ .\\
In both cases off the optimal damping, the absorption peaks are identical, but the half-absorption bandwidth is wider in the case of overdamping in contrast to underdamping.

\section{Decomposing reactance}
\label{sec:Quantifying}
\begin{figure*}[!htbp]
	\centering
	\includegraphics[width=1\textwidth]{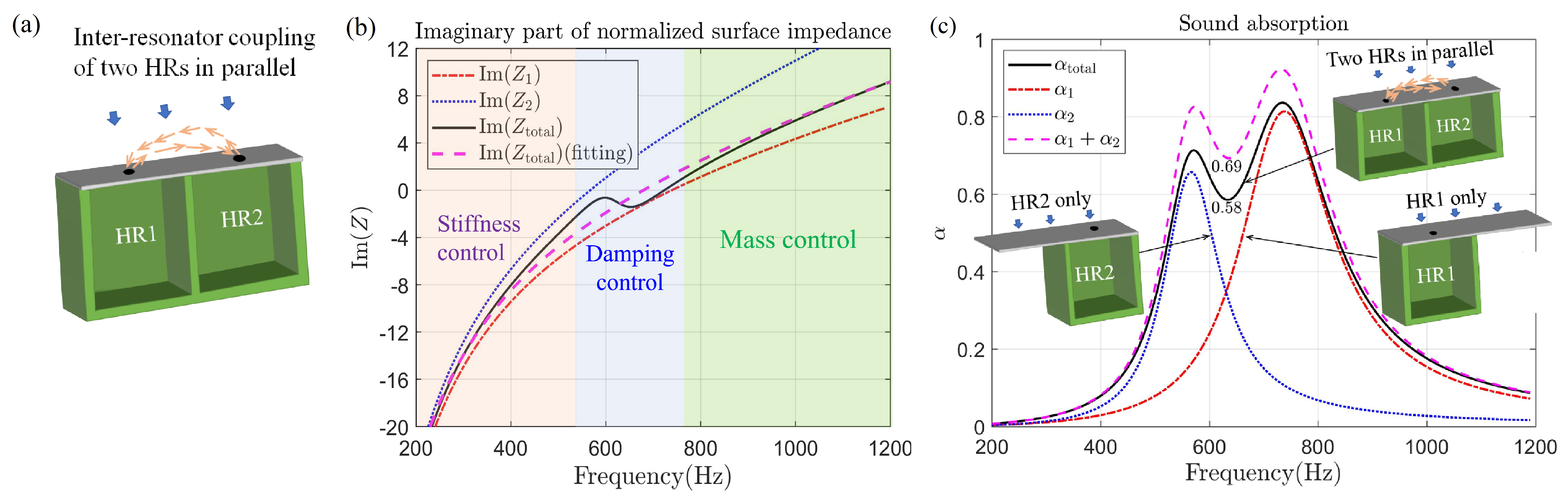}
	\caption{(a) Schematic of the inter-resonator coupling of two HRs in parallel. (b) $\mathrm{Im}(Z_1)$, $\mathrm{Im}(Z_2)$ and $\mathrm{Im}(Z_{\mathrm{total}})$ are normalized surface reactance of two individual resonators and them in parallel, respectively. $\mathrm{Im}(Z_{\mathrm{total}}){\mathrm{(fitting)}}$ is a least-mean-square fit for $\mathrm{Im}(Z_{\mathrm{total}})$ by Eq.~\eqref{impedance}. (c) $\alpha_{\mathrm{total}}$ is the total sound absorption of two resonators in parallel. $\alpha_1$ and $\alpha_2$ in red and blue dotted lines are the absorption coefficients derived from individual resonators with surface impedances $Z_1$ and $Z_2$, respectively. $\alpha_1+\alpha_2$ in pink dash line is the direct summation of $\alpha_1$ and $\alpha_2$.
 }
\label{Fig3}
\end{figure*}

Parallel resonator arrays have been extensively studied. When their absorption coefficient curves have been acquired through simulations or experiments, it is fascinating to determine the mass, spring, and damping based on the resonance peak for each resonator.
A parallel configuration of two HRs is shown in Fig.~\ref{Fig3}(a) in order to obtain the specific values of mass, spring, and damping, and thus better understand the physical mechanisms underlying the coupling effects.
The normalized input surface impedance $Z_\mathrm{total}$ from the top interface of an HR can be expressed as 
\begin{equation}\label{impedance}
   \begin{aligned}
	Z_\mathrm{total}  = \bar{\eta}_{\mathrm{HR}} + \mathrm{i}(\bar{M}_{\mathrm{HR}}\bar{\omega}-\bar{K}_{\mathrm{HR}}/\bar{\omega}),
  \end{aligned}
\end{equation}
where the mass term of the HR is given by $\bar{M}_{\mathrm{HR}}=M/(\rho_0 A_{\mathrm{n}} \sigma D)=l_{\mathrm{n}}/(\sigma D)$ with normalized angular frequency $\bar{\omega} = \omega D/c_0$ and resonator depth $D$. The resonance frequency can be calculated using the expression $\sqrt{\bar{K}_{\mathrm{HR}}/\bar{M}_{\mathrm{HR}}}c_0/(2\pi D)$.

Figure~\ref{Fig3}(b) depicts the normalized surface reactances of two individual HRs and their parallel configuration. A least-mean-square fit for $\mathrm{Im}(Z_{\mathrm{total}})$ using Eq.~\eqref{impedance} is represented by the purple dashed line. Similarly, $\mathrm{Im}(Z_1)$ and $\mathrm{Im}(Z_2)$ can also be fitted for the two HRs, i.e.,
\begin{equation}\label{impedanceZ12}
  \begin{aligned}
	Z_1 &= 1.35 + \mathrm{i}(54\bar{\omega}-0.97/\bar{\omega}),\\
       Z_2  &= 1.68 + \mathrm{i}(90\bar{\omega}-0.96/\bar{\omega}).
    \end{aligned}
\end{equation}
The expression of $\bar{M}_{\mathrm{HR}}$ reveals that the masses of these two HRs are amplified by a factor of $1/\sigma$, as demonstrated in Eq.~\eqref{eq:alpha_important_prime}. In contrast, the mass of the cavity alone, which possesses a depth of one-quarter wavelength, is expressed as $1/3$ as per Eq.~\eqref{eq:Taylor} in which $l_{\mathrm{coil}}$ reduces to $D$ in the absence of a coiling structure, assuming conformity in the form of $\bar{\omega}$. Consequently, the ratio of the HR mass to that of the cavity alone denotes the mass amplification factor, given as $3l_{\mathrm{n}}/(\sigma D)$. Remarkably, this factor represents a significant mass augmentation, i.e., 162 and 270 times greater than their respective pure cavities in mass, corresponding to $\lambda/648$ and $\lambda/1080$ respectively. This insightful observation concerning HRs further validates that sub-wavelength is a common phenomenon in acoustics, despite some claimed innovations of deep sub-wavelength behaviors, such as $\lambda/527$ and $\lambda/35$ in references~\cite{donda2019extreme,assouar2018acoustic,lemoult2013wave}.

Furthermore, the entire frequency band can be divided into three distinct regions, as shown in Fig.~\ref{Fig3}(b). The allocation of mass and stiffness control domains aligns intuitively with their opposite signs in Eq.~\eqref{impedance}. 
For instance, in the low-frequency range, the reactance exhibits negative values and is inversely proportional to frequency, thereby aligning with the stiffness term. Conversely, in the high-frequency range, the reactance becomes positive and is proportional to frequency, which matches the properties of the mass term. However, defining the damping control domain requires further consideration. In the case of two HRs in parallel, the total input impedance is defined as $Z=1/(0.5/Z_1+0.5/Z_2)$, where $Z_1 = \eta_1+\mathrm{i} X_1$ and $Z_2 = \eta_2+\mathrm{i} X_2$ with $X_1$ and $X_2$ the respective reactances. 
Specifically, for the frequency that is positioned approximately midway between the two peaks as shown in Fig.~\ref{Fig3}(c), a unique relationship emerges as $X_1=-X_2$. Consequently, the total input impedance can be rewritten and expressed as 
\begin{equation}\label{impedanceZ_para}
  \begin{aligned}
	Z = 2[\eta_1\eta_2+X_1^2+\mathrm{i}(\eta_1-\eta_2)X_1]/(\eta_1+\eta_2).
  \end{aligned}
\end{equation}
It is worth noting that the reactance component is transformed into damping by $X_1^2$, which represents a large new damping as it surpasses the original damping, i.e., $|X_1| > \eta_1,\eta_2$ as commonly observed in practical HRs. When combined with the term $\eta_1-\eta_2$, which attenuates the impact of the original reactance, this configuration prominently establishes a damping control domain.

\begin{table}[!tb] \vspace{0cm}
	\centering
	\caption{The equivalent damping $\bar{\eta}_{\mathrm{HR}}$, mass $\bar{M}_{\mathrm{HR}}$ and stiffness $\bar{K}_{\mathrm{HR}}$ for the two HRs and inter-resonator coupling from 592 Hz to 667 Hz.}
	\label{tab:MSD1} \vspace{0.2cm}
	\begin{tabular} {cccc}
		\hline
		& HR1 & HR2 & HR1+HR2 \\ [0.01cm] \hline
		$\bar{\eta}_{\mathrm{HR}}$ & $1.36$ & $1.70$ & 	\\
		$\bar{M}_{\mathrm{HR}}$ & $54$ & $90$ & $-40$	\\
		$\bar{K}_{\mathrm{HR}}$ & $0.96$ & $0.95$ & $-0.39$
		\\ [0.01cm] \hline
	\end{tabular} 
\end{table} \vspace{-0.0cm}
The sound absorption of these two HRs in parallel is obtained in Fig.~\ref{Fig3}(c), where optimization is performed by setting the objective frequency range as [500 Hz, 800 Hz]. The individual sound absorption coefficients under the optimal setup are depicted by the red dash dotted line and the blue dashed line, respectively.
The normalized oscillatory parameters of damping $\bar{\eta}_{\mathrm{HR}}$, mass $\bar{M}_{\mathrm{HR}}$, and stiffness $\bar{K}_{\mathrm{HR}}$, as listed in Table~\ref{tab:MSD1}, are utilized to assess the sound absorption performance of the two HRs with a narrow frequency range from 592 Hz to 667 Hz. The labels `HR1' and `HR2' refer to the equivalent oscillatory parameters of the uncoupled resonators, and the label `HR1+HR2' represents the equivalent oscillatory parameters of the two resonators in parallel. Both individual resonators exhibit positive values for their respective fitting parameters. However, when these two resonators are connected in parallel, it is observed that the equivalent mass and stiffness appear negative, i.e., $\bar{M}_{\mathrm{HR}}=-40$ and $\bar{K}_{\mathrm{HR}}=-0.39$. 


It should be noted that the presence of negative stiffness within this frequency range is meaningless in physical sense. 
In the damping controlling domain, as shown in Fig.~\ref{Fig3}(b), the reactance displays negative values which, in principle, should correspond to a positive stiffness. However, within this local narrow frequency range, there is an abrupt reversal in the curve's slope, which diverges from the prevailing trend represented by the solid black line. 
In other words, it is insufficient to consider only the negative equivalent mass or stiffness when evaluating sound absorption properties. In fact, one should also consider the variations of these two equivalent parameters with respect to frequency, i.e., their frequency derivatives~\cite{bobrovnitskii2014effective}.

\begin{figure*}[!htbp]
	\centering
	\includegraphics[width=1\textwidth]{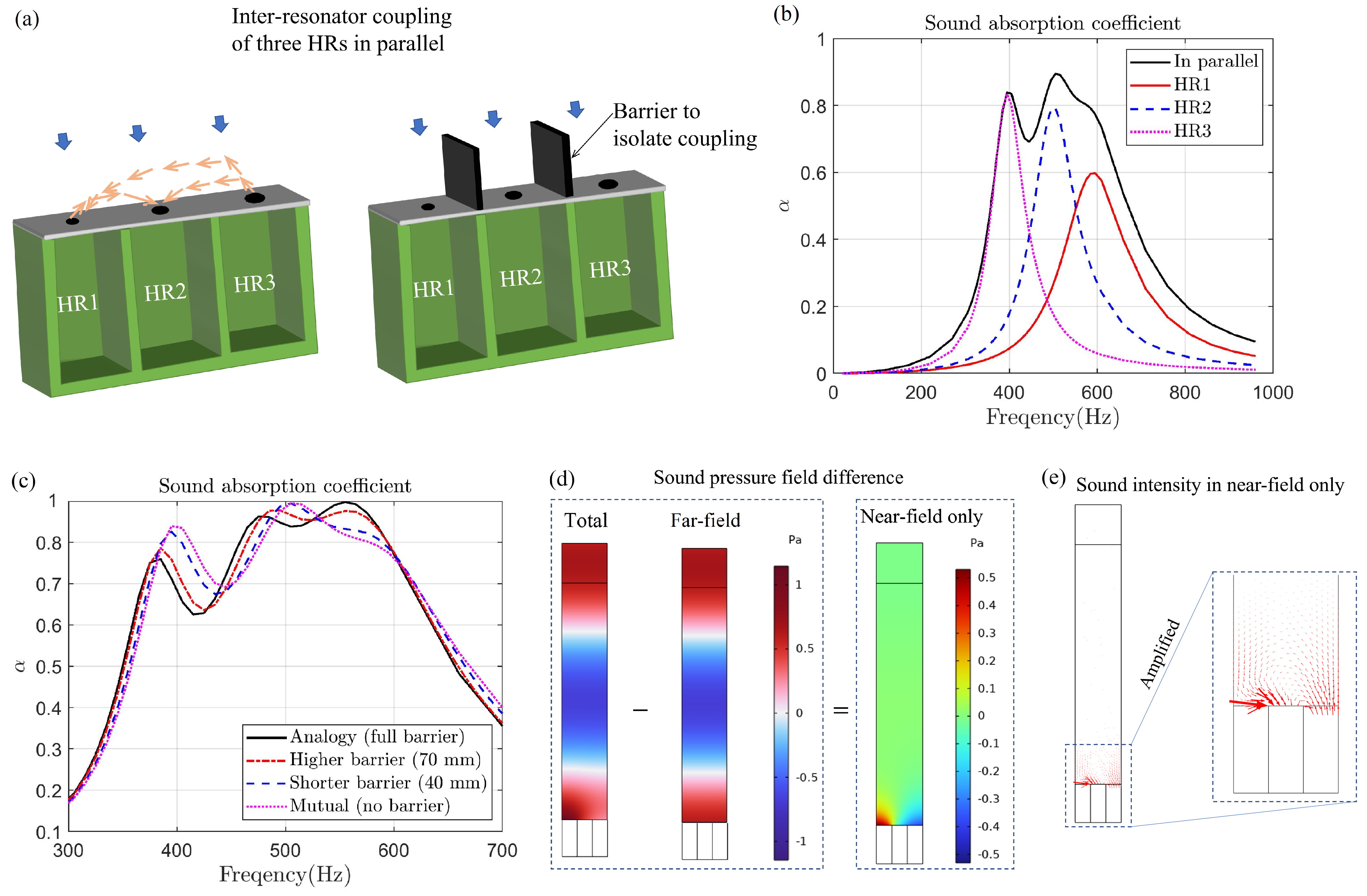}
	\caption{(a) Schematic of the inter-resonator coupling of three HRs in parallel. (b) Comparison of absorption performance among the individual resonators and the three HRs in parallel. (c) The influence of inter-resonator coupling on sound absorption through barriers of different heights to eliminate the near-field interaction. (d) The sound pressure in the near field is obtained by removing the far-field sound pressure from the total sound field. (e) The distribution of sound intensity flows in the near field illustrates the interaction among the three HRs.}
    \label{Fig2}
\end{figure*}

Many researchers~\cite{fang2006ultrasonic,kaina2015negative,kadic20193d,fleury2013extraordinary,cummer2016controlling} have reported a so-called performance enhancement of two HRs in parallel when compared to a single resonator. However, this comparison is indeed unfair in that the volume of the two HRs in parallel is twice that of a single resonator, as shown in Fig.~\ref{Fig3}(c). This volume doubling, which results in what they claimed as `negative stiffness', inherently leads to a half reduction in stiffness dominated by volume, as indicated by Eq.~\eqref{eq:Helm}. Hence, the concept of `negative stiffness' should not be ascribed to interactions or alternative interpretations; instead, it arises from the inconsistency in comparison baseline, which is misleading.
To provide a more accurate assessment involving stiffness, one should compare the sound absorption of the HRs in parallel with the summed absorption of the two individual resonators. In this scenario, the total volumes, i.e., stiffness, are held identical for a fair comparison. 

As shown in Fig.~\ref{Fig3}(c), the combined sound absorption performance of the HRs in parallel at the intersection point of the two HRs' individual absorption curves, in terms of absorption coefficient $\alpha_{\mathrm{total}}=0.58$, is actually inferior to the sum of absorption coefficients of the two individual HRs, i.e., $\alpha_1+\alpha_2=0.69$. Note that within the narrow frequency band between the two absorption peaks, one resonator with the lower resonance frequency acts more as a mass, while the other more as a spring. We speculate that a virtual resonator formed between the two inter-connected resonators occupies part of the volume (stiffness) out of the total system but does not contribute to the sound absorption in far field, as will be shown in Figs.~\ref{Fig2}(d) and~\ref{Fig2}(e). This finding indicates that the parallel coupling of the HRs essentially fails to enhance the overall sound absorption, let alone introduce any innovative impact related to `negative stiffness'. 
The deterioration in sound absorption can also be explained by the excessive overdamping resulting from reactance transformation, as has been shown in Eq.~\eqref{impedanceZ_para}.
Therefore, if someone claimed to have obtained mathematically negative stiffness with only passive strategies or without energy exchange with other physics fields, most likely it is merely a play on words. 

\section{Inter-resonator coupling}
\label{sec:coupling}
Extensive research efforts have been devoted to the study of resonator arrays with non-uniform impedance distributions.
Figure~\ref{Fig2}(a) provides an example of a parallel assembly of three HRs with the same resonant cavity yet different necks.
In Fig.~\ref{Fig2}(b), three curves of sound absorption coefficient are shown for each individual HR acting on their combined area (i.e., the same target area), along with the total sound absorption coefficient of the three HRs arranged in parallel and also acting on the combined area.
These sound absorption coefficients were calculated through simulations of finite element method using the Comsol software package. 

The interaction between the three HRs results in a mutual coupling of the resonator array, as has been discussed in \cite{shen2021acoustic}. To analyze the inter-resonator coupling, one approach is to introduce barriers of varying heights to isolate the resonators, as illustrated in Fig.~\ref{Fig2}(a). Figure~\ref{Fig2}(c) shows that the coupling effects becomes more pronounced as the barriers decrease in height, in that the resonance peak is gradually enhanced in amplitude and the peak frequency is shifted to higher frequencies, approaching the case without the barriers.
By subtracting the far-field sound pressure from the total sound pressure field of the HR array, we obtain the sound pressure map showing the near-field coupling in Fig.~\ref{Fig2}(d).
Following a similar procedure, Fig.~\ref{Fig2}(e) presents the sound intensity in the near field, depicting the flows of acoustic energy and the inter-resonator coupling effects.

\end{document}